\documentclass[prl,showpacs,showkeys,nofootinbib,twocolumn,floatfix]{revtex4}

\usepackage{graphicx}  
\usepackage{amsmath}
\usepackage{bm}  
\usepackage{ulem}
\usepackage{amssymb}
\usepackage{comment}
\usepackage{lineno}

\newcommand{\bea}{\begin{eqnarray}}
\newcommand{\eea}{\end{eqnarray}}
\newcommand{\beq}{\begin{equation}}
\newcommand{\eeq}{\end{equation}}

\begin{document}

\title{
Production of $\bm{X(3872)}$ at High Multiplicity
}

\author{Eric Braaten}
\email{braaten.1@osu.edu}
\affiliation{Department of Physics,
         The Ohio State University, Columbus, OH\ 43210, USA}

\author{Li-Ping He}
\email{he.1011@buckeyemail.osu.edu}
\affiliation{Department of Physics,
         The Ohio State University, Columbus, OH\ 43210, USA}

\author{Kevin Ingles}
\email{ingles.27@buckeyemail.osu.edu}
\affiliation{Department of Physics,
         The Ohio State University, Columbus, OH\ 43210, USA}

\author{Jun Jiang}
\email{jiangjun87@sdu.edu.cn}
\affiliation{School of Physics, Shandong University, Jinan, Shandong 250100, China}

\date{\today}

\begin{abstract}
The dependence of the production of the $X(3872)$ meson 
on the hadron multiplicity  in $pp$ collisions has been  used as
evidence against $X$ being a charm-meson  molecule.  The argument
is based in part on the incorrect assumption that the cross section for the breakup of  $X$ 
by scattering with comovers can be approximated by  a geometric cross section 
inversely proportional to the binding energy of $X$. 
The breakup cross section should instead be approximated
 by the probability-weighted sum of the cross sections for the scattering of comoving pions 
 from the charm-meson constituents of $X$, which is insensitive to the binding energy.
A simple modification of the comover interaction model gives excellent fits to
the data from the LHCb collaboration on the multiplicity dependence of the production of $X$ and $\psi(2S)$
using parameters compatible with $X$ being a loosely bound charm-meson  molecule.
\end{abstract}

\smallskip
\maketitle

{\bf Introduction.}
Since the unexpected discovery 
of the $X(3872)$ meson (also known as $\chi_{c1}(3872)$)  in 2003  \cite{Choi:2003ue},
dozens of other exotic heavy hadrons not predicted by the quark model have been discovered \cite{Ali:2017jda,Olsen:2017bmm,Karliner:2017qhf,Brambilla:2019esw}.
They present a major challenge to our understanding of QCD.
The nature of  $X(3872)$ ($X$ for short) is  a particularly important issue,
because it remains the exotic heavy hadron for which the most detailed experimental information is available.
The $X$ was discovered in the decay mode $J/\psi\,\pi^+\pi^-$, and it has since been observed in  6 other decay modes.
Its quantum numbers were determined  in 2013 to be $J^{PC}=1^{++}$ \cite{Aaij:2013zoa}.
The LHCb collaboration recently made the  most precise measurements of its mass $M_X$ 
and the first measurements of its decay width \cite{Aaij:2020xjx,Aaij:2020qga}.
The difference between $M_X$ and the $D^{*0}\bar{D}^0$ threshold is $\varepsilon_X=-0.07\pm 0.12$~MeV.
This implies an upper bound  on the binding energy of $X$: $|\varepsilon_X|<0.22$~MeV at 90\% confidence level.
 
The information $J^{PC}=1^{++}$ and 
$|\varepsilon_X|<0.22$~MeV is sufficient to conclude that $X$ must be
a loosely bound S-wave molecule with the particle content $(D^{*0}\bar{D}^0+D^0 \bar{D}^{*0})/\sqrt{2}$
and with universal properties determined by  $\varepsilon_X$  \cite{Braaten:2003he}.
The mean separation of its constituents is $r_X=(8\mu |\varepsilon_X|)^{-1/2}$,
where $\mu$ is the reduced mass of $D^{*0}\bar{D}^0$.
The upper bound $|\varepsilon_X|<0.22$~MeV implies  $r_X>4.8$~fm.
Thus this amazing hadron has a radius more than an order of magnitude larger than that of ordinary hadrons.
More relevant to the other exotic heavy hadrons is 
what $X$ would  have been if not for the fine-tuning of its mass to the $D^{*0}\bar{D}^0$ threshold.
The possibilities that have been proposed include
the P-wave charmonium state $\chi_{c1}(2P)$, 
an isospin-0 charm-meson molecule, and an isospin-1 compact tetraquark. 
In all these cases, the tuning of the mass to the $D^{*0}\bar{D^0}$ threshold  
produces resonant couplings to $D^{*0}\bar{D}^0$ and $D^0\bar{D}^{*0}$
that transforms $X$ into a loosely bound molecule of neutral charm mesons. 

Shortly after the discovery of $X$ in $B$-meson decays \cite{Choi:2003ue},
its existence was confirmed in $p\bar{p}$ collisions  \cite{Acosta:2003zx}.
The production of $X$ at a hadron collider can be resolved into two contributions:
 {\it prompt} production by strong interactions at the primary collision vertex 
and the  {\it $b$-decay} contribution from  weak decays of hadrons containing a bottom quark or antiquark  
at a displaced secondary vertex.
The behavior of these two contributions may provide evidence for the nature of $X$.
One significant difference is the hadronic environment in which $X$ is embedded.
In the decay of a $b$ hadron, at most a few additional hadrons emerge from the secondary vertex.
In prompt production at the LHC, hundreds of additional  hadrons may emerge from the primary vertex.
Collisions with comoving hadrons could  break $X$ up  into its charm-meson constituents
and thus decrease its prompt cross section.

The LHCb collaboration has studied the dependence on the hadron multiplicity of the
production of $X$ in $pp$ collisions  at the center-of-mass energy $\sqrt{s}=8$~TeV \cite{Aaij:2020hpf}. 
The charmonium state $\psi(2S)$ ($\psi^\prime$ for short) provides a convenient  benchmark, 
because  it also decays into $J/\psi\,\pi^+\pi^-$ and its mass is close to $M_X$.
The LHCb collaboration measured the ratio of the prompt production rates for $X$ and $\psi^\prime$ 
in the $J/\psi\,\pi^+\pi^-$ decay channel as functions of the number $N_\mathrm{tracks}$  of tracks in the vertex detector.
The prompt $X$-to-$\psi^\prime$ ratio decreases significantly with increasing  $N_\mathrm{tracks}$.

Esposito {\it et al.} have used the comover interaction (CI) model
to calculate  the dependence of the prompt $X$-to-$\psi^\prime$ ratio
on the charged-particle multiplicity $N_\mathrm{ch}$ \cite{Esposito:2020ywk}.
Their result if $X$ is a compact tetraquark is consistent with the LHCb  data,
while their result if $X$ is a molecule with a geometric cross section decreases much too rapidly with $N_\mathrm{ch}$. 
They concluded  that the LHCb data supports $X$ being a tetraquark
and strongly disfavors it being a  molecule. 
Their results if $X$ is a molecule were based in part on the incorrect assumption that its breakup reaction rate 
can be approximated by the  geometric cross section $\pi r_X^2$,  which is proportional to $1/|\varepsilon_X|$.
It should instead be approximated by the cross section
 for scattering  from the charm-meson constituents of $X$, which is  insensitive to  $\varepsilon_X$.
We show below that a simple modification of the CI model provides excellent fits to 
the LHCb data on the multiplicity dependence of $X$ and $\psi^\prime$ production
with parameters compatible with $X$  being a  loosely bound charm-meson molecule.

~

{\bf Comover Interaction Model.}
The CI model  was developed to describe the suppression of charmonium states in
relativistic $p$-nucleus and nucleus-nucleus  collisions by taking into account final-state interactions 
with comoving hadrons created by the collision \cite{Capella:1996va,Gavin:1996yd,Kharzeev:1996yx}.
Ferreiro used the CI model   \cite{Ferreiro:2014bia} to describe the
suppression of $\psi^\prime$ relative to $J/\psi$ 
 in $d$-Au and $p$-Pb collisions at RHIC \cite{Adare:2013ezl,Abelev:2014zpa,Arnaldi:2014kta}.
Ferreiro and Lansberg developed a more elaborate version of the CI model \cite{Ferreiro:2018wbd} 
to describe the suppression of  $\Upsilon(2S)$  and $\Upsilon(3S)$ 
relative to  $\Upsilon$ in $p$-Pb collisions at LHC \cite{Chatrchyan:2013nza,Aaboud:2017cif}. 
A modified version of their model was applied  by Esposito {\it et al.}\   
to the production of $X$ in $pp$ collisions \cite{Esposito:2020ywk}.

In the CI model,  the survival probability of a $c\bar{c}$ or  $b\bar{b}$ meson $\mathcal{Q}$
in $pp$ collisions is  \cite{Armesto:1997sa} 
\beq
S_\mathcal{Q} =  
\exp\left( - \frac{ \langle v \sigma_\mathcal{Q} \rangle\, dN/dy}{\sigma_{pp}}
 \log  \frac{dN/dy}{N_{pp}} \right) ,
\label{SX-Nch}
\eeq
where $dN/dy$ is the light-hadron multiplicity per unit range of rapidity
and $ \langle v\sigma_\mathcal{Q}\rangle$ is the  reaction rate
for the breakup of $\mathcal{Q}$ averaged over comovers. 
The nondiffractive cross section $\sigma_{pp}(s)$ depends on the center-of-mass energy $\sqrt{s}$, 
while $N_{pp}(s,y)$ may also depend on the rapidity $y$.
$N_{pp}$ is the multiplicity below which the effects of comovers are negligible:  $S_\mathcal{Q}=1$ if $dN/dy<N_{pp}$.
The estimates  for $\sigma_{pp}$  in Ref.~\cite{Esposito:2020ywk} 
are 63~mb  at $\sqrt{s}=7$~TeV and 70~mb at 13~TeV. 
A  logarithmic interpolation in $s$ gives $\sigma_{pp}=65$~mb at $\sqrt{s} = 8$~TeV. 
The range of pseudorapidity for the LHCb spectrometer is $2.0<\eta<4.8$.
An estimate of $N_{pp}$ in that region can be obtained by multiplying the mean charged-particle multiplicity
for the LHCb detector \cite{Aaij:2014pza}
by 3/2 to take into account neutral particles and then dividing by $\Delta y=2.8$, which gives $N_{pp} \approx 6$.

In the  CI model, the comovers are usually assumed to be either  
pions with mass $m_\pi\approx 140$~MeV or  massless gluons.
In Ref.~\cite{Ferreiro:2018wbd}, the momentum distribution of the comovers in the $\mathcal{Q}$  rest frame 
was assumed to be a Bose-Einstein distribution in the 2-dimensional transverse plane with an effective temperature $T_\mathrm{eff}$.
In Ref.~\cite{Esposito:2020ywk},  it was assumed to be a 3-dimensional Bose-Einstein distribution. 
Ref.~\cite{Ferreiro:2018wbd} introduced a  simplistic model for the  breakup cross section $\sigma_\mathcal{Q}$
as a function of the comover  energy $E_\pi$: 
$\pi r_\mathcal{Q}^2(1-E^\mathrm{thr}_\mathcal{Q}/E_\pi)^n$,
where $E^\mathrm{thr}_\mathcal{Q}$ is the threshold energy   for the breakup of $\mathcal{Q}$.
In  Ref.~\cite{Esposito:2020ywk}, that same model was used instead for the  breakup reaction rate $v  \sigma_\mathcal{Q}$.
In Ref.~\cite{Ferreiro:2018wbd}, $T_\mathrm{eff}$ was determined by fitting data on the 
the suppression of $\Upsilon(2S)$  and $\Upsilon(3S)$ in $p$-Pb and Pb-Pb collisions.
The fitted value of $T_\mathrm{eff}$ is approximately  linear  in $n$ 
between $\tfrac12$ and 2, and its extrapolation to $n=0$ is roughly 100~MeV.
For $n=1$, the effective temperature is $T_\mathrm{eff}=(250\pm 50)$~MeV.
These same values of $n$ and $T_\mathrm{eff}$ were used in Ref.~\cite{Esposito:2020ywk}. 

~

\begin{figure}[hbt]
\includegraphics*[width=0.92\linewidth]{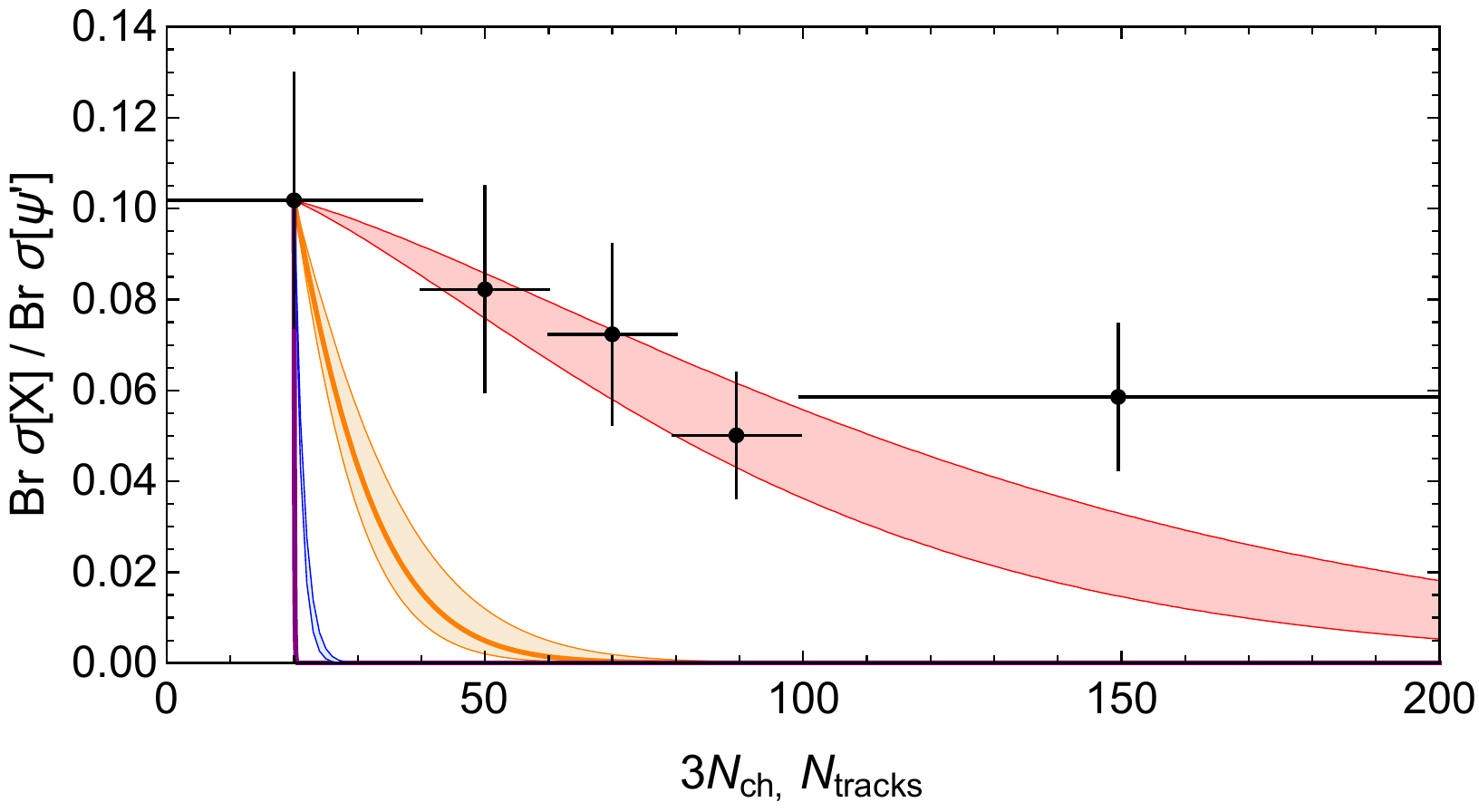} 
\caption{
Prompt $X$-to-$\psi^\prime$ ratio as a function of the multiplicity $N_\mathrm{ch}$.
The preliminary LHCb data from  Ref.~\cite{LHCb:2019obz} is shown in bins of $N_\mathrm{tracks}$,
which is identified with $3 N_\mathrm{ch}$.
The two  higher  error bands  are for a compact tetraquark
from  Ref.~\cite{Esposito:2020ywk} (red band, near the data)  
and using their value of $\langle v\sigma_X\rangle$ (orange band).
The two lower error bands  are for a charm-meson molecule 
from Ref.~\cite{Esposito:2020ywk} (blue band)  
and using their value of $\langle v\sigma_X\rangle$ (purple band, almost vertical).
}
\label{fig:Sratio-Esposito}
\end{figure}

{\bf Analysis of Ref.~\cite{Esposito:2020ywk}.}
In Ref.~\cite{Esposito:2020ywk}, their results for the  prompt $X$-to-$\psi^\prime$ ratio 
were compared with preliminary LHCb data \cite{LHCb:2019obz}.
The theoretical results were normalized to the first LHCb data point at  $N_\mathrm{tracks}=20$.
As shown in Fig.~\ref{fig:Sratio-Esposito},
their narrow error band for a molecule 
decreases precipitously to almost 0 near  $N_\mathrm{tracks} = 25$,
while their error band for a tetraquark gives a good fit to the LHCb data 
in the next three bins of $N_\mathrm{tracks}$, which extend from 40 to 100.  

It is implied  in Ref.~\cite{Esposito:2020ywk} that their error bands
follow from inserting their breakup reaction rates  $\langle v\sigma_\mathcal{Q} \rangle$
into the ratio $S_X/S_{\psi^\prime}$ of the survival probabilities given by  Eq.~\eqref{SX-Nch}.
The values of $\langle  v\sigma_\mathcal{Q}\rangle$ in Ref.~\cite{Esposito:2020ywk} are
$5.15\pm 0.84$~mb for $\psi^\prime$,  $11.61\pm 1.69$~mb for $X$ if it is a tetraquark,
and $1197\pm 171$~mb for $X$ if it is a molecule with $|\varepsilon_X|=116$~keV.
The prescription used to obtain these values was not specified. 
The resulting error bands are shown in Fig.~\ref{fig:Sratio-Esposito}.
The  error band using their value of
$\langle v\sigma_X \rangle$ if $X$ is a tetraquark decreases almost exponentially to 0, 
and it lies well below the LHCb data even in the second bin of $N_\mathrm{tracks}$.
Thus the error bands  in Ref.~\cite{Esposito:2020ywk} 
must  be determined by physics not captured by the survival probability in Eq.~\eqref{SX-Nch}.

~

{\bf $\bm{\pi X}$ Breakup Reaction Rates.}
Cross sections for low-energy $\pi X$ scattering can be calculated using
a nonrelativistic effective field theory for charm mesons and pions  called XEFT \cite{Fleming:2007rp}.
It provides a systematically improvable description of the sector of QCD consisting of 
$D^*\bar{D}$, $D\bar{D}^*$, $D\bar{D}\pi$, or $X$ with total energy near the $D^*\bar{D}$ threshold \cite{Fleming:2007rp}
and also the sector consisting of 
$D^*\bar{D}^*$, $D^*\bar{D}\pi$, $D\bar{D}^*\pi$, $D\bar{D}\pi\pi$, or $X\pi$
near the $D^*\bar{D}^*$ threshold  \cite{Braaten:2010mg}. 
A Galilean-invariant formulation of XEFT 
that exploits the approximate conservation of mass  in the transitions $D^*\leftrightarrow D\pi$
was introduced in Ref.~\cite{Braaten:2015tga} and further developed in Ref.~\cite{Braaten:2020nmc}.
Galilean invariance  guarantees that cross sections are the same in all Galilean frames, 
and it reduces the number of Feynman diagrams by requiring 
conservation of the total number of $\pi$, $D^*$, $\bar{D}^*$, and $X$ mesons.

\begin{figure}[hbt]
\includegraphics*[width=0.92\linewidth]{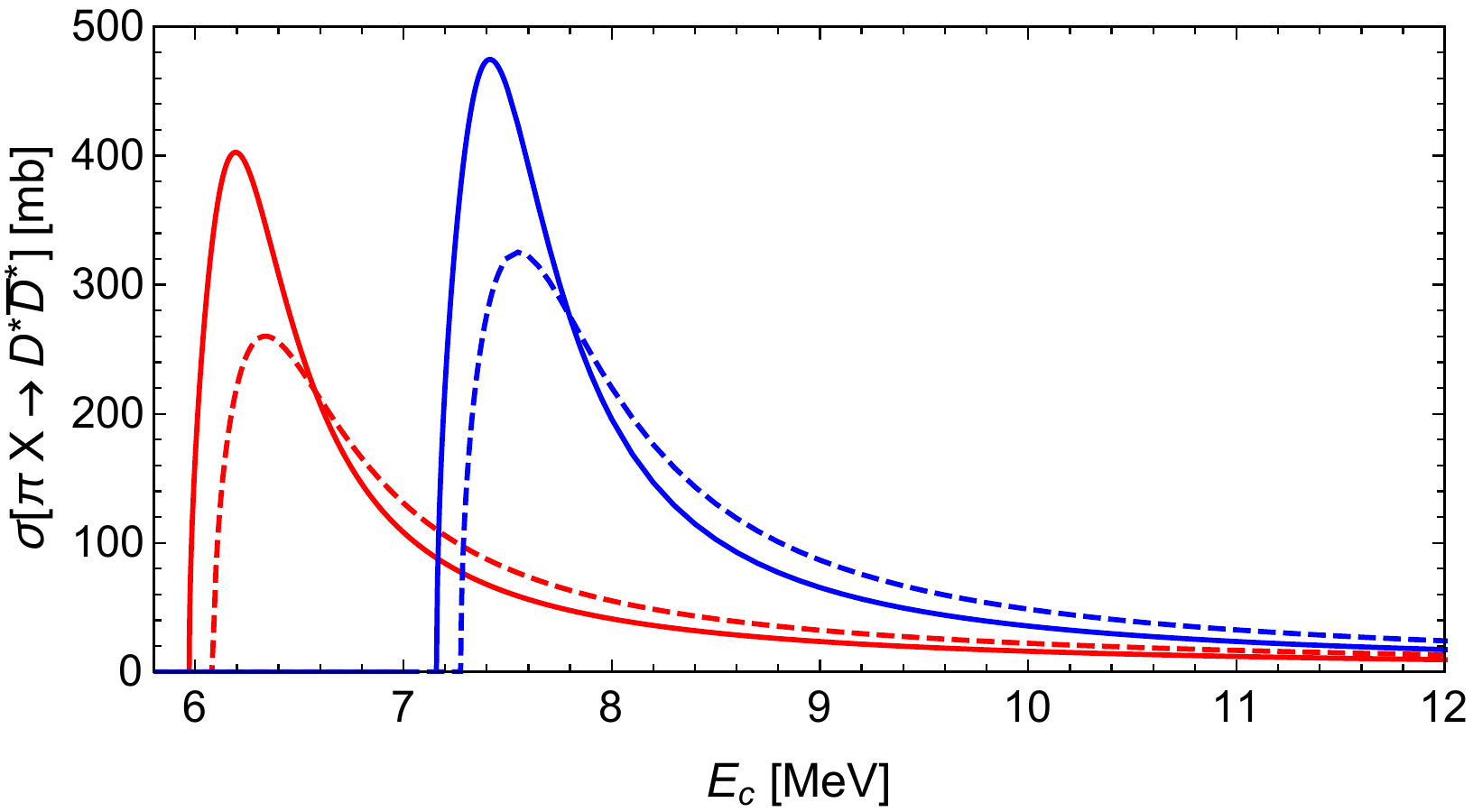} 
\caption{Breakup cross sections for $\pi^+X\to D^{*+}\bar{D}^0$ (red curves with lower threshold)
and $\pi^0X\to D^{*0}\bar{D}^0$ (blue curves with higher threshold)
as functions of the collision energy.
The binding energy of $X$ is 116~keV (solid curves) and  232~keV (dashed curves).
 }
\label{fig:sigmapi+/-X}
\end{figure}

The breakup cross section for  $\pi^+ X\to D^{*+}\bar{D}^{*0}$
was first calculated in Ref.~\cite{Braaten:2010mg} in the CM frame using original  XEFT. 
The cross sections for $\pi^+ X\to D^{*+}\bar{D}^{*0}$ and $\pi^0 X\to D^{*0}\bar{D}^{*0}$
are calculated  using Galilean-invariant XEFT in Ref.~\cite{piXscattering}. 
In Fig.~\ref{fig:sigmapi+/-X}, the cross sections are shown as functions of the collision energy $E_c$, 
which is the total kinetic energy in the CM frame.
They  have dramatic peaks near their thresholds, with peak values comparable to 
the geometric cross section $\pi r_X^2$, which is 1200~mb if  $\varepsilon_X = 116$~keV.
In the limit $\varepsilon_X\to 0$, the cross section for $\pi^+ X\to D^{*+} \bar{D}^{*0}$
approaches a delta function in $E_c$ at  $(\mu_{\pi X}/\mu_\pi)\delta_{0+}$,
where $\delta_{0+}= 5.9$~MeV is the $D^{*+}$-to-$D^0\pi^+$ energy difference and 
$\mu_{\pi X}$ and $\mu_\pi$ are the reduced masses for $\pi X$ and $\pi D$.
The energy-weighted integral of the cross section reduces in the limit  to
\beq
\int \!\!dE_c\, E_c\, \sigma[ \pi^+ X  \!\!\to\!\! D^{*+} \bar D^{*0}]  \longrightarrow 
\frac{2 \pi \sqrt2\,  \mu_{\pi X}^2 \delta_{0+}^{3/2} g^2}{\mu_\pi^{1/2} (2\sqrt{m_\pi} f_\pi)^2}   ,
\label{intEsigma:pi+X}
\eeq
where $g/(2\sqrt{m_\pi}f_\pi)$ is the $D^{*0}$-to-$D^0\pi^0$ coupling constant.
This is the integral required to calculate the contribution  
to $\langle v\sigma_X\rangle$ from a 3-dimensional Bose-Einstein distribution of pions.
The corresponding integral for $\pi^0 X\to D^{*0}\bar{D}^{*0}$ is obtained by replacing 
$\delta_{0+}$ by the $D^{*0}$-to-$D^0\pi^0$ energy difference $\delta_{00}= 7.0$~MeV.
Their contribution to $\langle v\sigma_X\rangle$ decreases from  0.2 to 0.04 to 0.02~mb
as $T_\mathrm{eff}$ increases from 100 to 200 to 300~MeV.

When the $\pi X$ collision energy is  well above the resonance region in Fig.~\ref{fig:sigmapi+/-X},
the pion can scatter off an individual constituent of $X$
and this will necessarily break up the  bound state. 
The constituents of $X$ are $D^{*0}$ and $\bar{D}^0$ with probability 1/2
and $D^0$ and $\bar{D}^{*0}$ with probability 1/2.
The total $\pi X$ breakup cross section can be approximated by the weighted
sum of $\pi D$ and $\pi D^*$ cross sections:
\beq
\sigma^\mathrm{incl}  [\pi X] \approx
\tfrac12 \Big( \sigma \big[\pi D^0\big] +  \sigma \big[\pi \bar D^0\big]
+ \sigma\big[\pi D^{*0} \big] + \sigma\big[\pi \bar D^{*0} \big] \Big).
\label{sigmapiXbreakup} 
\eeq
A  sufficient  condition for the validity of this approximation is that $E_c$  is well above the resonance region 
shown in Fig.~\ref{fig:sigmapi+/-X}.

For nonrelativistic collision energies,
the largest cross sections are those allowed in Galilean-invariant XEFT.
The specific final states from $\pi X$ scattering taken into account by Eq.~\eqref{sigmapiXbreakup} 
are $D^*\bar{D}\pi$ and $D\bar{D}^*\pi$ with at least one neutral charm meson.
The cross sections for $\pi D^0\to\pi D$ and $\pi D^{*0}\to\pi D^*$ are calculated  in Ref.~\cite{piXscattering}.
In the region  $\delta_{00}\ll E_c\ll m_\pi$, they are approximately constant.  
The total $\pi X$ breakup cross section  using Eq.~\eqref{sigmapiXbreakup} is
\beq
\sigma^\mathrm{incl}  [\pi X] \approx
\frac{4  \mu_\pi^2 (\mu_\pi^2 + \mu_{\pi *}^2)g^4}{\pi (2\sqrt{m_\pi} f_\pi)^4} ,
\label{sigmainclpiX:hi}
\eeq
where $\mu_{\pi *}$ is the  $\pi D^*$ reduced mass.
An over-estimate of the contribution of this region to $\langle v\sigma_X \rangle$
can be obtained by integrating over the range $\delta_{00}<E_c<m_\pi$.
This estimate decreases from 0.05 to 0.02 to 0.01~mb
as $T_\mathrm{eff}$ increases from 100 to 200 to 300~MeV.

For relativistic collision energies of order $m_\pi$ and larger, XEFT is not applicable.
In  Ref.~\cite{Cho:2013rpa}, a hadron scattering model was used to calculate
the contribution to $\langle v\sigma_X\rangle$ from the reactions 
$\pi X \to D^*\bar{D}^*$ in a thermal gas of hadrons. 
Their result decreases from 0.5 to 0.2~mb as  the temperature $T$ increases from 100 to 200~MeV.
In Ref.~\cite{Lin:2000jp}, a hadron scattering model was used to calculate the $\pi D$ and $\pi D^*$ 
reaction rates in a thermal gas of hadrons.
The structure of hadrons was taken into account by using a form factor with cutoff momentum $\Lambda$. 
For $\Lambda = \infty$, the  estimate of $\langle v\sigma_X\rangle$ using Eq.~\eqref{sigmapiXbreakup} 
increases from 25 to 37~mb as $T$ increases from 100 to 200~MeV,
while for $\Lambda = 1$~GeV, $\langle v\sigma_X\rangle \approx 15$~mb almost independent of $T$.

~

\begin{figure*}[hbt]
\includegraphics*[width=0.46\linewidth]{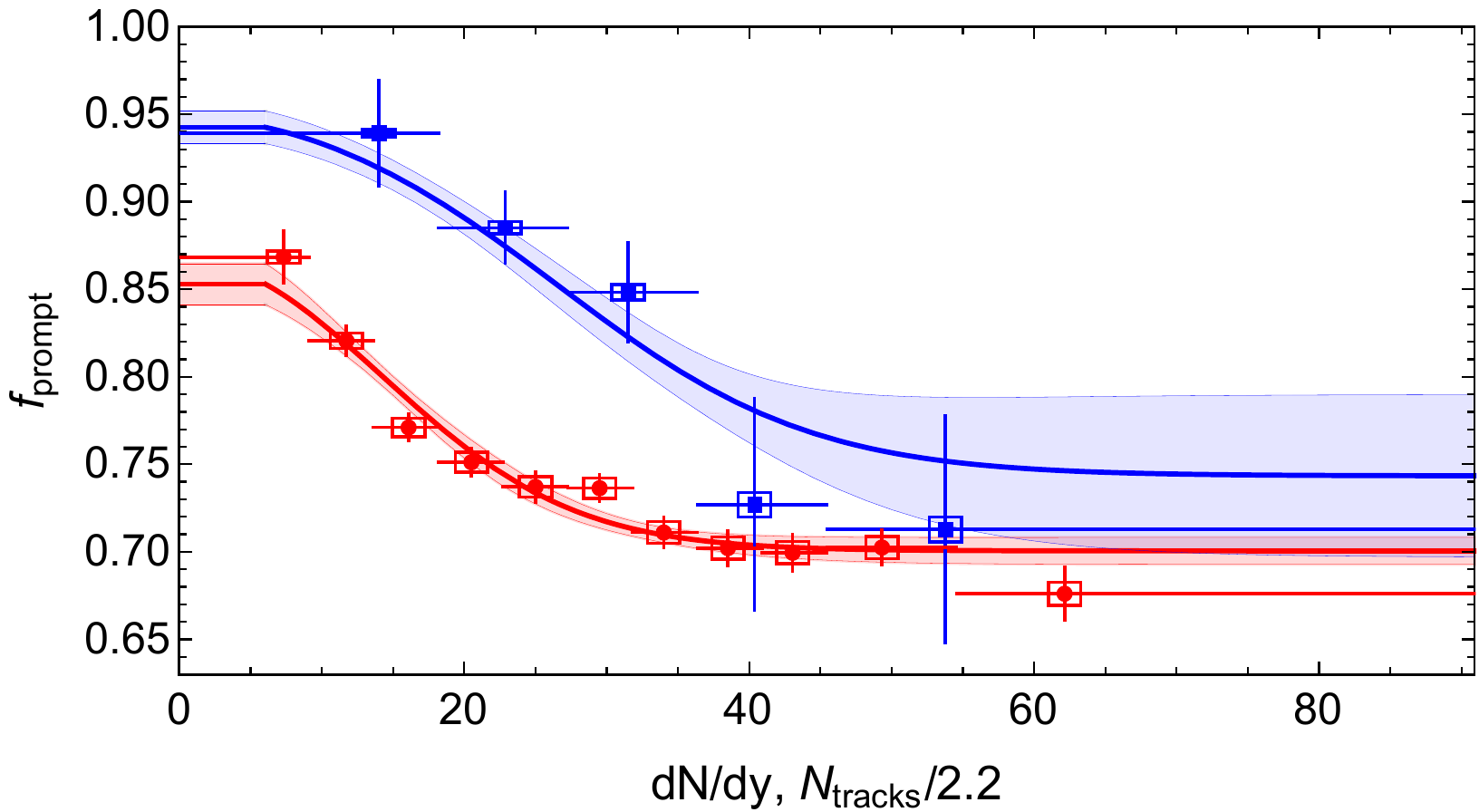} 
\hspace{1cm}
\includegraphics*[width=0.45\linewidth]{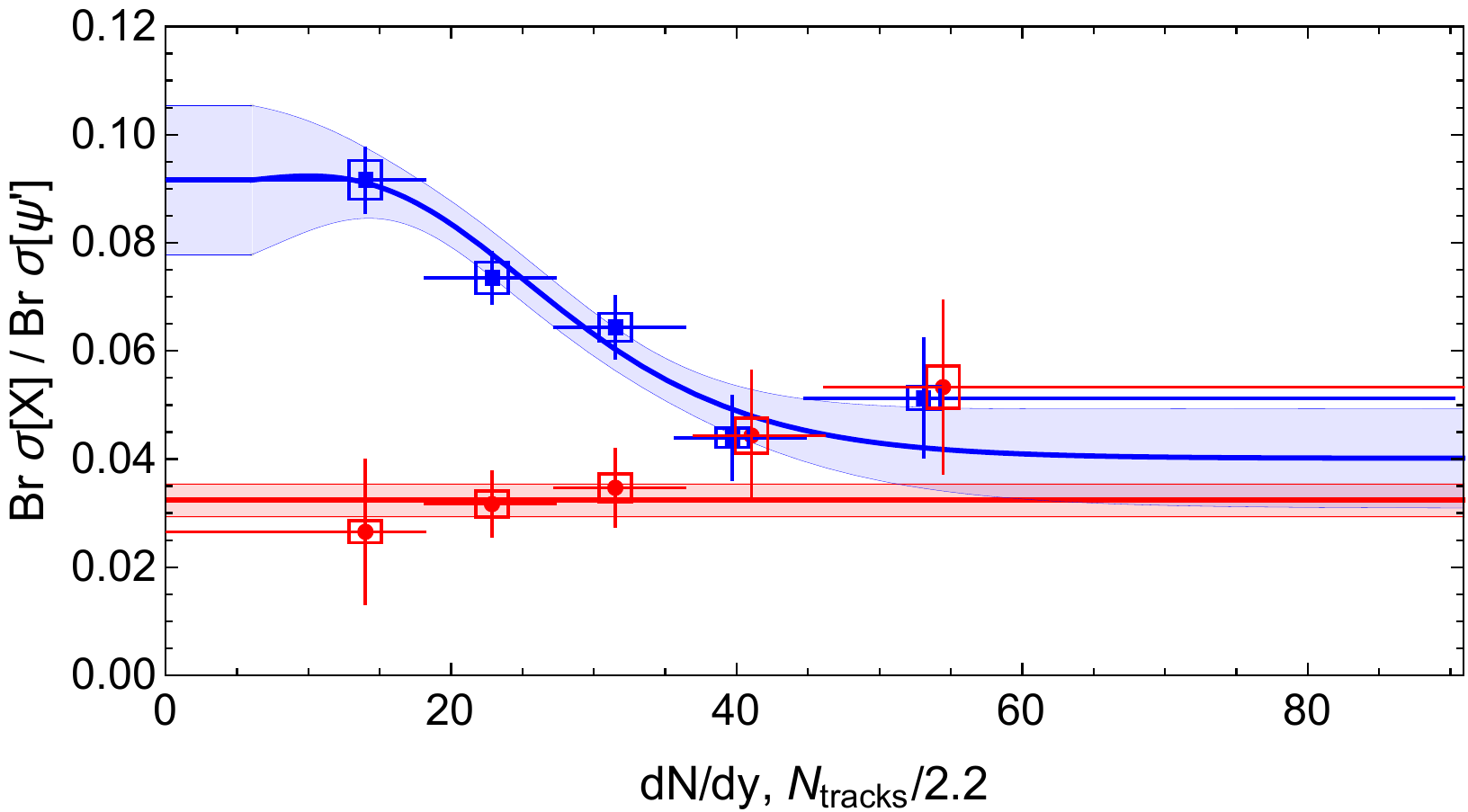} 
\caption{
Prompt fractions (left panel) and  $X$-to-$\psi^\prime$ production ratios (right panel)
as functions of the multiplicity $dN/dy$.
The LHCb data in Ref.~\cite{Aaij:2020hpf} for the prompt $X$  fraction (blue squares),  
the prompt $\psi^\prime$ fraction (red dots),
the prompt $X$-to-$\psi^\prime$ ratio (blue squares)  
and the $b$-decay  $X$-to-$\psi^\prime$ ratio (red dots) are shown in bins of $N_\mathrm{tracks}/2.2$.
The curves and their error bands are from a global fit to the LHCb data.
 }
\label{fig:fprompt}
\end{figure*}

{\bf Analysis of LHCb Data.}
The LHCb data in Ref.~\cite{Aaij:2020hpf}
consists of the prompt fractions for both $X$ and $\psi^\prime$ 
and the $X$-to-$\psi^\prime$ ratios for both prompt and $b$-decay production in Fig.~\ref{fig:fprompt}.
Since the pseudorapidity range of the LHCb vertex detector is $1.6<\eta<4.9$  \cite{Alves:2008zz}, 
the multiplicity $dN/dy$ can  be approximated by $\tfrac32 (N_\mathrm{tracks}/3.3)$.
The prompt fraction $f_\mathrm{prompt}$ for  $\psi^\prime$ is about 87\% in the first bin of  $N_\mathrm{tracks}$.
It first decreases as $N_\mathrm{tracks}$ increases, but then it appears to level off at about 70\%.
This behavior is incompatible with the assumption 
that the prompt cross section is proportional to the survival probability $S_{\psi^\prime}$
given by Eq.~\eqref{SX-Nch}. That assumption requires $f_\mathrm{prompt}$
to  decrease almost exponentially to 0 as  $N_\mathrm{tracks}$ increases.

A possible interpretation of the LHCb data on the prompt $\psi^\prime$ fraction in Fig.~\ref{fig:fprompt}
is that the prompt cross section has two components:
one  independent of   $dN/dy$ and  the other proportional to  $S_{\psi^\prime}$.
The two components could arise from the phase-space structure of the $pp$ collisions.
Prompt $\psi^\prime$’s created at a space-time point 
and with a momentum that  puts them out of reach of most of the comoving pions
give a contribution to the cross section that does not depend on $dN/dy$.
The remaining prompt $\psi^\prime$’s  are broken up with the probability $1-S_{\psi^\prime}$,
so their contribution to the prompt cross section is proportional to $S_{\psi^\prime}$.

This interpretation motivates a simple modification of the  CI model.
We denote the fraction of the prompt $\mathcal{Q}$  mesons
out of reach of comoving pions by $f_{\mathrm{out},\mathcal{Q}}$
and their contribution to the prompt cross section by $\sigma_{\mathrm{out},\mathcal{Q}}$.
The prompt cross section can be expressed as
\beq
\sigma_\mathrm{prompt}[\mathcal{Q}] =
\big[1 + (1/f_{\mathrm{out},\mathcal{Q}} - 1) S_\mathcal{Q} \big] \sigma_{\mathrm{out},\mathcal{Q}},
\label{sigma-prompt}
\eeq
which depends on $dN/dy$  through $S_\mathcal{Q}$.
We assume the $b$-decay cross section $\sigma_{b\,\mathrm{decay},\mathcal{Q}}$ does not depend on 
$dN/dy$.
The prompt fraction for $\mathcal{Q}$ is
\beq
f_\mathrm{prompt}[\mathcal{Q}] =
\frac{1 + (1/f_{\mathrm{out},\mathcal{Q}} - 1) S_\mathcal{Q}} 
{1 + 1/F_{\mathrm{out},\mathcal{Q}} + (1/f_{\mathrm{out},\mathcal{Q}} - 1) S_\mathcal{Q}} ,
\label{f-prompt}
\eeq
where $F_{\mathrm{out},\mathcal{Q}}=\sigma_{\mathrm{out},\mathcal{Q}}/\sigma_{b\,\mathrm{decay},\mathcal{Q}}$.
The  prompt $X$-to-$\psi^\prime$ ratio is
\beq
\frac{\mathrm{Br}\, \sigma_\mathrm{prompt}[X]}
{\mathrm{Br}\, \sigma_\mathrm{prompt}[\psi^\prime]}\,
= N_{X/\psi^\prime}
\frac{1 + (1/f_{\mathrm{out},X} - 1) S_X} 
{1 +  (1/f_{\mathrm{out},\psi^\prime} - 1) S_{\psi^\prime}} ,
\label{ratio:X/psi’-prompt}
\eeq
where $N_{X/\psi^\prime}$ is the product of $\sigma_{\mathrm{out},X}/\sigma_{\mathrm{out},\psi^\prime}$ 
and the ratio of the branching fractions into $J/\psi\,\pi^+\pi^-$.
The $b$-decay $X$-to-$\psi^\prime$ ratio  is
\beq
\frac{\mathrm{Br}\, \sigma_{b\,\mathrm{decay}}[X]}
{\mathrm{Br}\, \sigma_{b\,\mathrm{decay}}[\psi^\prime]}\,
 = N_{X/\psi^\prime} \frac{F_{\mathrm{out},\psi^\prime} }{F_{\mathrm{out},X} }  . 
\label{ratio:X/psi’-bdecay}
\eeq

We  have carried out a global fit to the LHCb data by minimizing the $\chi^2$ for the 26 data points in Fig.~\ref{fig:fprompt}
with respect to the 5 adjustable parameters in
Eqs.~\eqref{f-prompt}-\eqref{ratio:X/psi’-bdecay}
and the two breakup reaction rates $\langle v\sigma_X\rangle$  and $\langle v\sigma_{\psi^\prime}\rangle$.
The statistical and correlated errors were added in quadrature.
The resulting fits  are shown in Fig.~\ref{fig:fprompt}.
The error bands correspond to an increase of $\chi^2$  by less than 1.
The quality of the fits is very good with $\chi^2/\mathrm{dof}=0.99$.
The fit to the  $b$-decay $X$-to-$\psi^\prime$ ratio could be improved 
by adding a parameter that allows $\sigma_{\mathrm{out},\psi^\prime}$ or $\sigma_{b\,\mathrm{decay},X}$
 to increase linearly with $dN/dy$.
The  fractions of the prompt cross sections out of reach of comoving pions  are
$f_{\mathrm{out},\psi^\prime}=0.40\pm 0.03$ and $f_{\mathrm{out},X}=0.18\pm 0.04$. 
The ratios of the prompt and $b$-decay cross sections at large $dN/dy$ are
$F_{\mathrm{out},\psi^\prime}=2.3\pm 0.1$ and $F_{\mathrm{out},X}=2.9\pm 0.7$.  
The breakup reaction rates are $\langle v \sigma_{\psi^\prime}\rangle=3.9\pm 0.8$~mb
and $\langle v\sigma_X \rangle=2.6\pm 0.7$~mb. 
The prefactor in Eqs.~\eqref{ratio:X/psi’-prompt} and \eqref{ratio:X/psi’-bdecay} is $N_{X/\psi^\prime} = 0.04 \pm 0.01$. 

The fitted value of $\langle v\sigma_{\psi^\prime}\rangle$ is about 1$\sigma$ smaller than 
the  value in Ref.~\cite{Esposito:2020ywk}. 
It is about 5$\sigma$ smaller than 
the  value in Ref.~\cite{Esposito:2020ywk} if $X$ is a tetraquark.
The fitted value of $\langle v\sigma_X \rangle$ is about 4 times larger  than the contribution from  
$\pi X \to D^*\bar{D}^*$ in a thermal gas of hadrons with $T=100$~MeV in Ref.~\cite{Cho:2013rpa}.
The fitted value of $\langle v\sigma_X \rangle$  is less than 1/4 the total breakup reaction rate from Ref.~\cite{Lin:2000jp}. 
This could be attributed to a failure of the
3-dimensional Bose-Einstein distribution as a model for comoving pions.
A momentum distribution that is isotropic in the two transverse dimensions and the longitudinal dimension
seems implausible.

~

{\bf Outlook.}
The LHCb data on the multiplicity dependence of the production of $X$ and $\psi^\prime$ in $pp$ collisions
is incompatible with the assumption that the prompt cross section is proportional to  
the survival probability in Eq.~\eqref{SX-Nch}.
However, as shown  in Fig.~\ref{fig:fprompt},
a good global fit can be obtained by adding the assumption that some fraction $f_{\mathrm{out},\mathcal{Q}}$
of the prompt $\mathcal{Q}$ cross section is out of reach of comoving pions.
A  microscopic description of $pp$ collisions in which these fractions could be calculated would be useful.

The quantum  numbers $J^{PC}=1^{++}$ and the upper bound $|\varepsilon_X|<0.22$~MeV imply that
$X$ must be a loosely bound S-wave molecule  of neutral charm mesons 
 with universal properties determined by $\varepsilon_X$.
Universality is a double-edged sword.
It allows definite statements about some properties of $X$, such as $r_X$
 and the $\pi X$ breakup reaction rate,
 but it also makes them insensitive to what $X$ would have been
if not for the fine-tuning of its mass to the $D^{*0}\bar{D}^0$ threshold.
$X$ could have been a more compact charmonium or molecule or tetraquark,
but it is transformed into a large neutral-charm-meson molecule
by its resonant interactions with $D^{*0}\bar{D}^0$ and $D^0\bar{D}^{*0}$.
Given the upper bound on $|\varepsilon_X|$, a model for $X$ as a compact hadron
should be interpreted as a fictitious hadron that  does not couple to the 
charm mesons at the nearby $D^{*0}\bar{D}^0$ threshold.
It may be an interesting exercise to rule out such a possibility using experimental data, 
but it  is already excluded by theoretical considerations.

The universal physics of a loosely bound S-wave molecule reveals a dramatic failure of the simplistic model 
in Refs.~\cite{Esposito:2020ywk} and \cite{Ferreiro:2018wbd} for 
$\langle v \sigma_\mathcal{Q}\rangle$ based on the geometric cross section $\pi r^2_\mathcal{Q}$. 
That model overestimates  $\langle v \sigma_X\rangle$  by orders of magnitude.
The breakup cross section $\sigma_X$ is comparable to  $\pi r^2_X$, 
which is proportional to $1/|\varepsilon_X|$, only at energies very close to the threshold as shown in Fig.~\ref{fig:sigmapi+/-X}.
At higher energies,  $\sigma_X$ is determined by the cross sections for scattering from the constituents of $X$ in Eq.~\eqref{sigmapiXbreakup},
so $\langle v \sigma_X\rangle$ is insensitive to $\varepsilon_X$.

Our fit to the LHCb data in Fig.~\ref{fig:fprompt}  may be
a step towards a  quantitative understanding of  the production of $X$ in high-energy hadron collisions.
In a hadron collision, once $X$ is broken up into charm mesons by 
the  collision with a comoving pion, the probability that one of the charm mesons will encounter another 
charm meson and that they will coalesce  into $X$ is extremely small.
An attempt to calculate the coalescence contribution to the production in $pp$ collisions of $X$ if it is 
a molecule was made in Ref.~\cite{Esposito:2020ywk}.
Coalescence can be much more important in $p$-nucleus and nucleus-nucleus collisions,
because the number of charm meson that are created is much larger.
The first observation of the production of $X$ in heavy-ion collisions by the CMS collaboration
indicated that the prompt $X$-to-$\psi^\prime$ ratio may be much larger 
in Pb-Pb collisions than in $pp$ collisions \cite{CMS:2019vma}.
Understanding  the production of $X$ in $p$-nucleus and nucleus-nucleus collisions 
even at the qualitative level remains a challenging open problem.

~

\begin{acknowledgments}
This work was supported in part by the U.S.\ Department of Energy under grant DE-SC0011726,
the National Natural Science Foundation of China under grant 11905112,
and the Natural Science Foundation of Shandong Province under grant ZR2019QA012.
We acknowledge useful communications with A.~Esposito and A.~Pilloni.
\end{acknowledgments}






\begin{thebibliography}{99}

\bibitem{Choi:2003ue} 
  S.K.~Choi {\it et al.} [Belle Collaboration],
Observation of a narrow charmonium-like state in exclusive $B^\pm \to K^\pm \pi^+ \pi^-  J/\psi$ decays,
  Phys.\ Rev.\ Lett.\  {\bf 91}, 262001 (2003)
  [hep-ex/0309032].
  
\bibitem{Ali:2017jda} 
  A.~Ali, J.S.~Lange and S.~Stone,
Exotics: Heavy Pentaquarks and Tetraquarks,
  Prog.\ Part.\ Nucl.\ Phys.\  {\bf 97}, 123 (2017)
  [arXiv:1706.00610].
  
\bibitem{Olsen:2017bmm} 
  S.L.~Olsen, T.~Skwarnicki and D.~Zieminska,
Nonstandard heavy mesons and baryons: Experimental evidence,
  Rev.\ Mod.\ Phys.\  {\bf 90}, 015003 (2018)
  [arXiv:1708.04012].

\bibitem{Karliner:2017qhf} 
  M.~Karliner, J.L.~Rosner and T.~Skwarnicki,
Multiquark States,
  Ann.\ Rev.\ Nucl.\ Part.\ Sci.\  {\bf 68}, 17 (2018)
  [arXiv:1711.10626].
    
\bibitem{Brambilla:2019esw}
N.~Brambilla, S.~Eidelman, C.~Hanhart, A.~Nefediev, C.P.~Shen, C.E.~Thomas, A.~Vairo and C.Z.~Yuan,
The $XYZ$ states: experimental and theoretical status and perspectives,
Phys.\ Rept.\ \textbf{873}, 1 (2020)
[arXiv:1907.07583].

\bibitem{Aaij:2013zoa} 
R.~Aaij {\it et al.} [LHCb Collaboration],
Determination of the $X(3872)$ meson quantum numbers,
  Phys.\ Rev.\ Lett.\  {\bf 110}, 222001 (2013)
  [arXiv:1302.6269].
  
\bibitem{Aaij:2020xjx}
R.~Aaij \textit{et al.} [LHCb],
Study of the $\psi_2(3823)$ and $\chi_{c1}(3872)$ states in $B^+ \rightarrow \left( J\psi\pi^+\pi^-\right)K^+$ decays,
JHEP \textbf{08}, 123 (2020)
[arXiv:2005.13422].

\bibitem{Aaij:2020qga}
R.~Aaij \textit{et al.} [LHCb],
Study of the lineshape of the $\chi_{c1}(3872)$ state,
Phys.\ Rev.\ D \textbf{102}, 092005 (2020)
[arXiv:2005.13419].

\bibitem{Braaten:2003he}
E.~Braaten and M.~Kusunoki,
Low-energy universality and the new charmonium resonance at 3870 MeV,
Phys.\ Rev.\ D \textbf{69}, 074005 (2004)
[hep-ph/0311147].

\bibitem{Acosta:2003zx}
D.~Acosta \textit{et al.} [CDF],
Observation of the narrow state $X(3872) \to J/\psi \pi^+ \pi^-$ in $\bar{p}p$ collisions at $\sqrt{s} = 1.96$ TeV,
Phys.\ Rev.\ Lett.\ \textbf{93}, 072001 (2004)
[arXiv:hep-ex/0312021].

\bibitem{Aaij:2020hpf}
R.~Aaij \textit{et al.} [LHCb],
Observation of multiplicity-dependent prompt $\chi_{c1}(3872)$ and $\psi(2S)$ production in $pp$ collisions,
[arXiv:2009.06619 [hep-ex]].

\bibitem{Esposito:2020ywk}
A.~Esposito, E.G.~Ferreiro, A.~Pilloni, A.D.~Polosa and C.A.~Salgado,
The nature of $X(3872)$ from high-multiplicity $pp$ collisions,
[arXiv:2006.15044 [hep-ph]].

\bibitem{Capella:1996va}
A.~Capella, A.~Kaidalov, A.~Kouider Akil and C.~Gerschel,
$J / \psi$ and $\psi^\prime$ suppression in heavy ion collisions,
Phys.\ Lett.\ B \textbf{393}, 431 (1997)
[arXiv:hep-ph/9607265].

\bibitem{Gavin:1996yd}
S.~Gavin and R.~Vogt,
Charmonium suppression by Comover scattering in Pb + Pb collisions,
Phys.\ Rev.\ Lett.\ \textbf{78}, 1006 (1997)
[arXiv:hep-ph/9606460].

\bibitem{Kharzeev:1996yx}
D.~Kharzeev, C.~Lourenco, M.~Nardi and H.~Satz,
A Quantitative analysis of charmonium suppression in nuclear collisions,
Z.\ Phys.\ C \textbf{74}, 307 (1997)
[arXiv:hep-ph/9612217].

\bibitem{Ferreiro:2014bia}
E.G.~Ferreiro,
Excited charmonium suppression in proton-nucleus collisions as a consequence of comovers,
Phys.\ Lett.\ B \textbf{749}, 98-103 (2015)
[arXiv:1411.0549].

\bibitem{Adare:2013ezl}
A.~Adare \textit{et al.} [PHENIX]
Nuclear Modification of $\psi^\prime$, $\chi_c$, and $J/\ensuremath{\psi}$ Production in d+Au Collisions at $\sqrt{s_{NN}}$=200  GeV,
Phys.\ Rev.\ Lett.\ \textbf{111},  202301 (2013)
[arXiv:1305.5516].

\bibitem{Abelev:2014zpa}
B.B.~Abelev \textit{et al.} [ALICE],
Suppression of $\psi$(2S) production in p-Pb collisions at $\sqrt{s_{\rm NN}}$ = 5.02 TeV,
JHEP \textbf{12}, 073 (2014)
[arXiv:1405.3796].

\bibitem{Arnaldi:2014kta}
R.~Arnaldi [ALICE],
Inclusive $\psi$(2S) production in p-Pb collisions with ALICE,
Nucl.\ Phys.\ A \textbf{931}, 628-632 (2014)
[arXiv:1407.7451].

\bibitem{Ferreiro:2018wbd}
E.G.~Ferreiro and J.P.~Lansberg,
Is bottomonium suppression in proton-nucleus and nucleus-nucleus collisions at LHC energies due to the same effects?,
JHEP \textbf{10}, 094 (2018)
[arXiv:1804.04474].

\bibitem{Chatrchyan:2013nza}
S.~Chatrchyan \textit{et al.} [CMS],
Event Activity Dependence of $\Upsilon(nS)$ Production in $\sqrt{s_{NN}}$=5.02 TeV pPb and $\sqrt{s}$=2.76 TeV pp Collisions,
JHEP \textbf{04}, 103 (2014)
[arXiv:1312.6300].

\bibitem{Aaboud:2017cif}
M.~Aaboud \textit{et al.} [ATLAS],
Measurement of quarkonium production in proton\textendash{}lead and proton\textendash{}proton collisions at $5.02~\mathrm {TeV}$ with the ATLAS detector,
Eur.\ Phys.\ J.\ C \textbf{78}, 171 (2018)
[arXiv:1709.03089].

\bibitem{Armesto:1997sa}
N.~Armesto and A.~Capella,
A Quantitative reanalysis of $J / \psi$ suppression in nuclear collisions,
Phys.\ Lett.\ B \textbf{430}, 23 (1998)
[arXiv:hep-ph/9705275].

\bibitem{Aaij:2014pza}
R.~Aaij \textit{et al.} [LHCb],
Measurement of charged particle multiplicities and densities in $pp$ collisions at $\sqrt{s}=7\;$TeV in the forward region,
Eur. Phys. J. C \textbf{74}, 2888 (2014)
[arXiv:1402.4430].

\bibitem{LHCb:2019obz}
LHCb collaboration,
Multiplicity-dependent modification of $\chi_{c1}(3872)$ and $\psi{(2S)}$ production in $pp$ collisions at $\sqrt{s}$= 8 TeV,
LHCb-CONF-2019-005.

\bibitem{Fleming:2007rp} 
  S.~Fleming, M.~Kusunoki, T.~Mehen and U.~van Kolck,
Pion interactions in the $X(3872)$,
  Phys.\ Rev.\ D {\bf 76}, 034006 (2007)
  [hep-ph/0703168].

\bibitem{Braaten:2010mg}
E.~Braaten, H.~Hammer and T.~Mehen,
Scattering of an Ultrasoft Pion and the $X(3872)$,
Phys.\ Rev.\ D \textbf{82}, 034018 (2010)
[arXiv:1005.1688].
  
\bibitem{Braaten:2015tga} 
  E.~Braaten,
Galilean-invariant effective field theory for the $X(3872)$,
  Phys.\ Rev.\ D {\bf 91},  114007 (2015)
  [arXiv:1503.04791].  

\bibitem{Braaten:2020nmc}
E.~Braaten, L.-P.~He and J.~Jiang,
Galilean-Invariant XEFT at Next-to-Leading Order,
[arXiv:2010.05801 [hep-ph]].

\bibitem{piXscattering}
E.~Braaten, L.-P.~He, K.~Ingles, and J.~Jiang, 
Scattering of Pions with $X(3872)$,
in preparation. 

\bibitem{Cho:2013rpa}
S.~Cho and S.H.~Lee,
Hadronic effects on the X(3872) meson abundance in heavy ion collisions,
Phys.\ Rev.\ C \textbf{88}, 054901 (2013)
[arXiv:1302.6381].

\bibitem{Lin:2000jp}
Z.w.~Lin, T.G.~Di and C.M.~Ko,
Charm meson scattering cross-sections by pion and rho meson,
Nucl.\ Phys.\ A \textbf{689}, 965 (2001)
[arXiv:nucl-th/0006086].

\bibitem{Alves:2008zz}
A.A.~Alves, Jr. \textit{et al.} [LHCb],
The LHCb Detector at the LHC,
JINST \textbf{3}, S08005 (2008).

\bibitem{CMS:2019vma}
 [CMS],
Evidence for $\chi_{c1}$(3872) in PbPb collisions and studies of its prompt production 
at $\sqrt{\smash[b]{s_{_{\mathrm{NN}}}}}=5.02$ TeV,
CMS-PAS-HIN-19-005.


  \end{thebibliography}
 \end{document}